\documentclass[manuscript]{acmart}

\setcopyright{none}
\renewcommand\footnotetextcopyrightpermission[1]{} % removes footnote with conference information in first column
\AtBeginDocument{%
  \providecommand\BibTeX{{%
    \normalfont B\kern-0.5em{\scshape i\kern-0.25em b}\kern-0.8em\TeX}}}

\setcopyright{acmcopyright}
\copyrightyear{2018}
\acmYear{2018}
\acmDOI{XXXXXXX.XXXXXXX}

\begin{document}

\title{Bridging Deliberative Democracy and Deployment of Societal-Scale Technology}

\author{Ned Cooper}
\email{edward.cooper@anu.edu.au}
\orcid{0000-0003-1834-279X}
\affiliation{%
  \institution{College of Engineering, Computing \& Cybernetics, The Australian National University}
  \city{Canberra}
  \country{Australia}
}

\maketitle
This position paper encourages the Human-Computer Interaction (HCI) community to focus on designing deliberative processes to inform and coordinate technology and policy design for large language models (LLMs)---a `societal-scale technology'. First, I propose a definition for societal-scale technology and locate LLMs within this definition. Next, I argue that existing processes to ensure the safety of LLMs are insufficient and do not give the systems democratic legitimacy. Instead, we require processes of deliberation amongst users and other stakeholders on questions about the safety of outputs and deployment contexts. This shift in AI safety research and practice will require the design of corporate and public policies that determine how to enact deliberation and the design of interfaces and technical features to translate the outcomes of deliberation into technical development processes. To conclude, I propose roles for the HCI community to ensure deliberative processes inform technology and policy design for LLMs and other societal-scale technology.

\section{Societal-Scale Technology}

In this position paper, I define `societal-scale technology' as an artefact or system that is:
\begin{itemize}
    \item developed based on interactions with society, or
    \item impacts society once deployed.
\end{itemize}
I define society in this position paper as groupings of people across political, economic, geographical, and cultural boundaries. For example, I consider LLMs to be societal-scale technology, as LLMs interact with training data that represent groupings of people across such boundaries, and, at least in recent deployment cases (\textit{e.g.}, the ChatGPT research preview), LLMs have impacted groups of people across such boundaries. Within such societies, ChatGPT has impacted direct stakeholders (those who interact directly with ChatGPT, such as school students) and indirect stakeholders (those who may or may not have interacted directly with ChatGPT but are affected by the interaction of direct stakeholders with ChatGPT, such as school teachers) \citep{friedman2019value}.

\section{Processes of Deliberation for LLMs}

While LLMs may interact with representations of societies during development in the form of training datasets, the group of people actively developing LLMs rarely, if ever, reflect the boundary-spanning nature of the societies in which LLMs are deployed. For example, the workforce of OpenAI is not (and could never be) as diverse as the societies in which they deployed ChatGPT. The values and preferences of those groups of people actively developing an LLM are reflected in any system they develop, regardless of the representation of diverse values and preferences in a training dataset---through the specification of system behaviour once deployed. In the case of ChatGPT, for example, this was achieved through content filters, among other mechanisms. The content filters specified specific values and preferences that OpenAI intended ChatGPT to reflect. However, I contend that the specification of values and preferences by one group of developers is insufficient to ensure the safety of LLMs and does not give such systems democratic legitimacy. A small group of expert developers cannot adequately foresee the safety risks of technology deployed across societies---composed of multiple professions and stakeholders---let alone speak for those stakeholders during development and deployment \citep{joyce2021toward}.

Instead, I encourage those considering safe development and deployment strategies for LLMs to review the emphasis of political science literature over recent decades on deliberation as the essence of democratic legitimacy \citep{fishkin1991democracy,curato2017twelve,knight1994aggregation}. Deliberative democracy involves an association of members who govern their affairs by deliberation among the members \citep{cohen2005deliberation}. In other words, deliberative democracy is about making collective decisions determined by those subject to the decisions: not only their preferences, interests, and choices but also their reasoning \citep{elster1998deliberative}. If algorithmic fairness is primarily a political task, as \citet{wong2020democratizing} argues, rather than solely a technical task, we must consider how to resolve issues politically rather than technically. In the societal-scale context of LLMs, instead of groups of developers within individual organisations resolving such political questions for us, I argue that we must design processes that provide people using a system (\textit{i.e.}, direct stakeholders) or people affected by a system (\textit{i.e.}, indirect stakeholders) with the opportunity to deliberate with others (including developers) on how the system functions, and in what contexts to deploy those systems.

\section{Roles for HCI in Deliberative AI}

OpenAI accompanied the release of ChatGPT with a request for feedback from users on the outputs of the system to improve its safety. As stated in the blog post announcing the release of ChatGPT:
\begin{quote}
    \textit{"We are interested in feedback regarding harmful outputs that could occur in real-world, non-adversarial conditions, as well as feedback that helps us uncover and understand novel risks and possible mitigations." } \cite{openai2022}
\end{quote}

Yet, the scope of feedback accepted through the ChatGPT interface is severely limited---to indications of approval or disapproval of an individual output of the system and some indication by the user of what an `ideal' answer might have been. This feedback process remains tightly controlled by those deploying the system and focuses on aligning outputs to individual user preferences. It does not facilitate deliberation \textit{amongst} stakeholders regarding the proper outputs of LLMs nor the appropriate deployment contexts for LLMs.

HCI researchers and practitioners are well-placed to build deliberative capacity into AI safety research and practice---the HCI community takes users' preferences seriously and interrogates the methods through which we elicit user preferences. To this end, I envisage three roles for the HCI community for the safe development and deployment of LLMs:
\begin{enumerate}
    \item Designing corporate and public policies that define criteria for and conditions of membership of associations for deliberation
    \item Developing platforms or interfaces that facilitate bidirectional communication among developers and members of associations, and multi-directional communication amongst the members of an association
    \item Designing processes for documentation of collective decisions, and research on how to link documented decisions to technical development processes (\textit{e.g.}, how could the processes of Reinforcement Learning from Human Feedback expand to include deliberation amongst stakeholders or to facilitate feedback from collectives of users?).
\end{enumerate}
\clearpage
\bibliographystyle{ACM-Reference-Format}
\bibliography{main.bib}

\end{document}